\documentclass[prc,showpacs,superscriptaddress]{revtex4} 
\usepackage{rotating}
\usepackage{graphicx}
\begin{document}

\title{The Upscattering of Ultracold Neutrons from the polymer [C$_{6}$H$_{12}$]$_n$}

\author{E. I. Sharapov}
\affiliation{Joint Institute for Nuclear Research, 141980, Dubna, Russia}
\author{C. L. Morris}
\thanks{Corresponding author}
\email{morris@lanl.gov}
\affiliation{Los Alamos National Laboratory, Los Alamos, NM 87544, USA}
\author{M. Makela}
\affiliation{Los Alamos National Laboratory, Los Alamos, NM 87544, USA}
\author{A. Saunders}
\affiliation{Los Alamos National Laboratory, Los Alamos, NM 87544, USA}
\author{Evan R. Adamek}
\affiliation{Department of Physics, Indiana University, Indiana 47405-7105 USA}
\author{L. J. Broussard}
\affiliation{Los Alamos National Laboratory, Los Alamos, NM 87544, USA}
\author{C. B. Cude-Woods}
\affiliation{Department of Physics, Indiana University, Indiana 47405-7105 USA}
\author{Deion E Fellers}
\affiliation{Los Alamos National Laboratory, Los Alamos, NM 87544, USA}
\author{Peter Geltenbort}
\affiliation{Institut Laue-Langevin, 38042 Grenoble Cedex 9, France}
\author{M. Hartl}
\affiliation{Los Alamos National Laboratory, Los Alamos, NM 87544, USA}
\author{S. I. Hasan}
\affiliation{Department of Physics and Astronomy, University of Kentucky, Lexington, Kentucky 40506, USA}
\author{K. P. Hickerson}
\affiliation{Kellogg Radiation Laboratory, California Institute of Technology, Pasadena, California 91125, USA.}
\author{G. Hogan}
\affiliation{Los Alamos National Laboratory, Los Alamos, NM 87544, USA}
\author{A. T. Holley}
\affiliation{Department of Physics, Indiana University, Indiana 47405-7105 USA}
\author{C. M. Lavelle}
\affiliation{Applied Physics Laboratory, The Johns Hopkins University, Baltimore, Maryland 21218 USA}
\author{Chen-Yu Liu}
\affiliation{Department of Physics,Indiana University, Indiana 47405-7105 USA}
\author{M. P. Mendenhall}
\affiliation{Kellogg Radiation Laboratory, California Institute of Technology, Pasadena, California 91125, USA.}
\author{J. Ortiz}
\affiliation{Los Alamos National Laboratory, Los Alamos, NM 87544, USA}
\author{R. W. Pattie Jr.}
\affiliation{Department of Physics, North Carolina State University, Raleigh, North Carolina 27695, USA}
\author{D. G. Phillips II}
\affiliation{Department of Physics, North Carolina State University, Raleigh, North Carolina 27695, USA}
\author{J. Ramsey}
\affiliation{Los Alamos National Laboratory, Los Alamos, NM 87544, USA}
\author{D. J. Salvat}
\affiliation{Department of Physics, Indiana University, Indiana 47405-7105 USA}
\author{S. J. Seestrom}
\affiliation{Los Alamos National Laboratory, Los Alamos, NM 87544, USA}
\author{E. Shaw}
\affiliation{Los Alamos National Laboratory, Los Alamos, NM 87544, USA}
\author{Sky Sjue}
\affiliation{Los Alamos National Laboratory, Los Alamos, NM 87544, USA}
\author{W. E. Sondheim}
\affiliation{Los Alamos National Laboratory, Los Alamos, NM 87544, USA}
\author{B. VornDick}
\affiliation{Department of Physics, North Carolina State University, Raleigh, North Carolina 27695, USA}
\author{Z. Wang}
\affiliation{Los Alamos National Laboratory, Los Alamos, NM 87544, USA}
\author{T. L. Womack}
\affiliation{Los Alamos National Laboratory, Los Alamos, NM 87544, USA}
\author{A. R. Young}
\affiliation{Department of Physics, North Carolina State University, Raleigh, North Carolina 27695, USA}
\author{B. A. Zeck}
\affiliation{Department of Physics, North Carolina State University, Raleigh, North Carolina 27695, USA}


\date{August 2, 2013}

\begin{abstract}

It is generally accepted that the main cause of ultracold neutron (UCN)  losses in storage traps  
is the upscattering to the thermal energy range by hydrogen adsorbed on the surface of the trap walls. 
However, the data on which this conclusion is based 
are poor and contradictory. Here, we report a measurement, performed at the Los Alamos National Laboratory UCN source, of 
the average energy of the flux of upscattered neutrons after the interaction of UCN with hydrogen 
bound in semicrystalline polymer PMP (tradename TPX),
[C$_{6}$H$_{12}$]$_n$. 
Our analysis, performed with the MCNP code based 
on the application of the neutron scattering law to 
UCN upscattered by bound hydrogen in semicrystalline polyethylene, 
[C$_{2}$H$_{4}$]$_n$, 
leads us to a flux average energy value of 26$\pm3$ meV in contradiction with 
previously reported experimental values of 10 to 13 meV 
 and  in agreement with the theoretical  
models of neutron heating implemented in the MCNP code. 
\end{abstract}
\vspace{1pc}
\pacs{ 68.49.-h, 78.70.Nx, 68.47.-b} 
\maketitle

UCN are neutrons with kinetic energy below a critical value of ∼ 100
neV (velocity 4.4 m/s). This corresponds to an average temperature of
≃1 mK,  hence the technical term `ultra cold`). 
The UCN neutrons  are totally reflected from material surfaces at all
angles of 
incidence and therefore can be confined in traps 
for time intervals of several hundred seconds $-$ comparable to the
neutron 
lifetime. Recent reviews 
\cite{Snow013,Lamor09} highlight the use of UCN in nuclear and particle physics, cosmology and gravity. In particular, an importance of study of the upscattering spectrum temperature and rates for neutron lifetime measurements is emphasized in \cite{Pick09}.
Authors of Ref. \cite{Golub04}  discuss using UCN upscattering techniques in solid state and surfaces studies 
with the view that the rather limited UCN intensity available at existing UCN sources will be overcome with more powerful 
next generation UCN sources. 
For all these studies a better understanding of UCN interactions, especially upscattering,  at surfaces of different materials 
continues to be of importance in a view of  existing inconsistencies in data, as detailed in \cite{Golub04}.
During the transport in neutron guides or the storage in the material traps, UCN can be excited above the critical energy by absorbing energy from thermal excitation of the surface 
materials and leave the confinement space. 
It is presently believed that the main reason for UCN heating in traps  is 
 inelastic scattering from the hydrogen molecules on the material surfaces. Indeed, by the nuclear reaction 
 analysis method \cite{Land77} with $^{15}$N beam impinging on the unbaked copper samples, the area density 
of the surface hydrogen was determined to be 2x10$^{16}$ H/cm$^2$ and the hydrogen containing layer was estimated to be 
3.0 nm  thick. \\

Hydrogen has one of the largest inelastic scattering
cross section of all elements. It is most convenient to study it in polyethylene,  
[C$_{2}$H$_{4}$]$_n$, in which the hydrogen inelastic cross section
increases  
as the inverse velocity law to the value of 2053$\pm$40 barn
\cite{Pok011} 
at a velocity of 4.0 m/s. 
Authors of Refs. \cite{Stoika78,Str78}, 
used a 100-$\mu$m thick polyethylene sample inside a UCN trap and
$\sim$4$\pi$ $^3$He  neutron detectors of a different gas pressure to 
 observe the upscattered neutron flux and have determined its average
 energy to be 10-13 meV 
(the corresponding velocity $\sim$1600 m/s). 
This is  about half of 
the value expected from the known phonon frequency spectrum of hydrogen in polyethylene, and has triggered 
further efforts to find possible channels of the UCN escaping from traps.  Authors of Ref. \cite{Muz99}, using the 
neutron activation method, searched for lower energy upscattering of UCN from a Be surface. A possibility of 
scattering in the range of 15 to 300 m/s was convincingly ruled out. Also, so called "weak heating"
with UCN energy changes $\simeq 10$ neV above the critical energy has been reported \cite{Str00}, 
with a probability estimated to be $\sim 10^{-7}$ per the collision with  copper, which is rather low value. 
In this report we study UCN upscattering into the thermal energy range in polymethylpentene (PMP), [C$_{6}$H$_{12}$]$_n$, which, due to
its low 
density (0.83 g/cm$^3$),  is a material  having
negative, practically non-reflective 
optical potential $U_{F}$ with the value even smaller than $U_{F}$ for
a  more often used polyethylene (PE),
[C$_{2}$H$_{4}$]$_n$. Our PMP (tradename TPX, Mitsui Chemicals,
Inc. for the atactic poly[4-methyl-1-pentene] material) was a 
tetragonal of Form I semicrystalline sample with XRD spectrum 
similar to that of the undrawn sample in \cite{He87} 
as was evidenced by our own X-ray
diffraction data.  \\

The measurements have been performed at the Los Alamos National Laboratory solid-deuterium ultracold neutron source 
driven by the 800 MeV, 5.8-$\mu$A average proton beam provided by the Los Alamos Neutron Science Center (LANSCE) linear accelerator.  
The source is described in details in a recent publication by Saunders et al. \cite{Andy013}. 
The scheme of the experimental geometry is shown in Fig. 1. 
A 7.62 cm diameter stainless steel UCN guide tube (1) was connected 
to the main UCN guide through the section 
containing a flange with a zirconium foil embedded in the 6 T field of  a
 superconducting solenoid magnet. This Zr foil separates the UCN source vacuum system from the external guide.
The magnetic field accelerates half of the neutrons above the foil’s
critical energy allowing the high field seaking beam to exit the UCN
source, reducing losses due to transmission through the foil.  
The 500 $\mu$m thick PMP sample (2) with the same diameter as the 
inside of the  tube was installed  at the end of the shown section. 
The density of the UCN beam in the tube  was $\simeq$1.0 UCN/cm$^3$, and the average 
velocity of the UCN flux spectrum was $\simeq$4 m/s.  
Two 5 cm diameter, 30 cm long, drift tube 3He neutron detectors (3)
and (4) were installed perpendicular to the
guide axis 
and symmetrically above and below it, 
as shown in Figure 1, to provide equal fluxes for both detectors.
The construction, gas filling 
and performance of detectors are described in \cite{Chris09}. The
partial $^3$He pressure was 1.8 bar in one detector and 0.2 bar in the other one. 
Data were accumlated for 300 sec under steady beam conditions.
The up-scattered neutron rates were significantly above the
background, which was mostly due to cosmogenic thermal neutrons. The
background associated with the proton beam was eliminated using time gates
on the analog to digital converters, to reject events during the beam pulses.
The neutron rates for the analysis were calculated by integrating the 
measured  pulse height spectra from both detectors. The ratio  
R = N$_{1.8}$/N$_{0.2}$ of integrated rates  was 
compared with calculations to deduce the average energy
$\langle$E$\rangle$  
of the upscattered flux. 
Data were also taken with a 6.3 mm thick polyethylene slab inserted  
between detectors and the end of the UCN guide  to 
thermalize completely the upscattered neutrons and to compare the
result 
with calculations. \\

In principle, if the scattered flux has Maxwellian shape, the flux average energy $\langle$E$\rangle$   can be easily estimated following the following argument. The efficiencies $\epsilon$ of the $^3$He detectors for neutron 
with energy $E$ crossing the tubes
along the same track $\ell$  are different
because $\epsilon = 1- \exp \{ -N(^3He) \ell \sigma(E)\}$ while the  Helium-3 atomic densities $N(^3He)$
 are 
different for the 1.8 bar and 0.2 bar detectors ($\sigma(E)$ is well
known cross section for the $^3$He(n,p)T reaction). We calculated efficiencies for several Maxwellian spectra with different 
$\langle$E$\rangle$ using the code MCNP5 \cite{Mcnp5} to follow all neutron trajectories crossing our detectors in the 
geometry of Fig. 1. A calculation of $R$ vs. $\left\langle E \right\rangle$ (shown in fig. 2)
provides then the average energy  of the Maxwellian upscattered spectrum using the
measured value of R. 
However, Maxwellian shape for the upscattered flux 
is not expected from the theory 
of neutron inelastic scattering by a bond hydrogen.

For UCN upscattering,  a simple analytical formula for 
the isotropic differential
cross section in 
 one-phonon incoherent approximation of the theory  
\begin{equation}
\frac{d\sigma_{up}}{dE} = \sigma_{\rm b} \sqrt\frac {E}{E_{i}} \left (e^{E/kT}-1\right
)^{-1}\frac {g(E)} {\rm A }e^{-2W} 
\end{equation}
demonstrates clearly that the upscattering spectrum is not Maxwellian.
This equation, written initially in terms of the neutron wave
vector variable, was obtained by Placzek and Van Hove \cite{Pla54}.   
Here $\sigma_{\rm b} = 4\pi {\rm b}^2$ is the cross section for bound
nuclei with the mass number ${\rm A }$, $E_{i}$ 
is the initial UCN energy, $E$ is the energy after upscattering,
$g(E)$ is the generalized 
(amplitude weighted) phonon density of states in the material under
study 
and the last exponent is the material- and temperature-dependent
Debye-Waller factor. At a temperature of 300K, 
multi-phonon contributions are also important 
\cite{Kita68}  
and a full upscattering spectrum is often   
calculated then with the use of the MCNP code and its models for 
the neutron scattering law S$(\alpha, \beta)$  \cite{MacFar010}    
($\alpha$ and $\beta$ 
are neutron reduced momentum and energy transfers) which takes some
account for multiphonon processes.
The issue of different theoretical approaches to multiphonon
scattering is of interest itself but we leave it out of the present
report 
because our experimental data can't distinguish between them.
A most recent model of the generalized density of states of hydrogen in PE 
is provided by Barrera et al. \cite{Barr06}. It was validated recently
by neutron inelastic measurements \cite{Lave013} 
for the high density PE. Using  these data and the deduced 
S$(\alpha,\beta)$  one of the
 authors (C.M.L.) created the 77K  and 293K scattering
 kernels, represented as tables of the double differential 
scattering cross sections for modeling the neutron  transport in PE. 
Although such data are absent for the polymer PMP, we believe 
that its neutron scattering law  is similar to the S$(\alpha,\beta)$
of PE and one can use PE as a model for UCN upscattering from PMP. 
Indeed, the known experimental infrared spectra are similar 
for PMP \cite{PMP013} and PE \cite{Krimm56} and the same is 
valid for numerous theoretical calculations, 
as referenced for example in \cite{Kum09}. Of course, the presence 
of the methyl groups in the side branches of the PMP molecular chain can 
additionally influence the low energy range of PMP phonon density of
states, not accessible in infrared and Raman   spectroscopy,
and we adress this issue in the concluding paragraph.    \\

Therefore, with the PE new kernels added to the MCNP thermal energy data 
we modeled the scattered flux from our PMP sample in the 
experimental geometry of Fig. 1 with an initial 
neutron beam  energy $E_{in} = 100$ neV. The results are shown in Fig. 3. The shape of the directly upscattered 
flux (black squares) is clearly not Maxwellian and  was found to be independent of the initial neutron energy 
and the placement of detectors. The spectrum of 
the fully moderated (by the additional 6.3-mm thick polyethylene slab
placed 
between the sample and detectors) flux is Maxwellian. 
Having proved these spectral shapes we modeled  
the $^3$He(n,p)T reaction  rates in the two detectors after the UCN
scattering   
using the 293K kernel. 
The results of modeling together with the  results of measurements
are presented in TABLE I  for comparison.\\

\begin{table}[ht]
\caption{Experimental results in comparison with the MCNP-modeling. $\rm{N(1.8)}$ and $\rm{N(0.2)}$ 
represent count rates taken during the 300 s run with detectors of the 1.8 bar and 0.2 bar $^3$He pressure. 
$\rm{R(exp)}$ and $\rm{R(mcnp)}$ represent their ratios. }
\label{tab:tb1}
\begin{tabular} {||c|c|c|c|c||}\hline\hline
\rule{0pt}{4ex}Neutron spectrum & $\rm{N(1.8)}$ &$\rm{N(0.2)}$ &$\rm{R(exp)}$ &$\rm{R(mcnp)}$\\
\hline
\rule{0pt}{4ex} Upscattered &71159 &14943 &4.76$\pm$ 0.07 & 4.67 \\
Moderated   &32360 &5264  &6.14$\pm$ 0.09 & 5.90
 \\
\hline
  &  & & &  \\
${\mathrm {DOUBLE\;RATIO}}$ &   &   & 0.77$\pm$0.02  &0.79 \\
\hline
\hline
\end{tabular}
\end{table}
As shown in TABLE I, the measurements on the PMP sample and the
MCNP modeling 
with the use of the 
PE 293K kernel based on 
theoretical S$(\alpha, \beta)$ scattering law agree rather well. 
In the modeling, the non-Maxwellian 
scattered spectrum for a thin sample has the average energy value of
26$\pm 3$ meV. 
Therefore we conclude 
that, after the initial UCN energy, the spectrum of neutrons,
scattered in one 
or only few interactions,  
has the expected non-Maxwellian shape and the average energy
$\langle$E$\rangle$=26$\pm 3$ meV. At the same time, the 
fully moderated flux spectrum has Maxwellian shape with the average
energy of 
53$\pm 4$ meV at room temperature. 
Our conclusion contradicts obviously with the value
$\langle$E$\rangle=10 -13$ meV 
\cite{Stoika78,Str78}, 
obtained in an essentially the same kind of the experiment 
although with a different $^3$He detector, with a PE sample  
and by the analysis  in a frame of Maxwellian approximation. 
This situation induced us to reanalyze data \cite{Stoika78,Str78}
using the MCNP 
code with the input 
describing geometry and conditions of the measurement \cite{Str78}. 
The result is shown in Fig. 4, 
where the full squares are experimental data of \cite{Str78} 
and curves 1 and 2 have
been 
obtained by the MCNP modeling 
with the scattering law S($\alpha,\beta$) kernels at 77K   and 293K 
to produce non-Maxwellian theoretical spectra with
$\langle$E$\rangle$=11 meV and $\langle$E$\rangle$=26 meV, correspondly.
 The curve 3 for the 293K Maxwellian
spectrum is shown for comparison. From agreement of experimental data  
with the curve 2 we
claim  the average energy of 26 meV in the PE experiment of Stoika 
and Strelkov \cite{Str78} $-$ the same average energy as in our PMP
experiment modeled with the neutron scattering law S$(\alpha, \beta)$ 
for PE.

This comparison of two experiments allows us to conclude that, 
as suggested in the text of
this report, the phonon densities of states in polyethylene and 
polymethylpentene  are similar. Finding possible low energy
distinctions between them  
would require 
dedicated UCN and thermal neutron inelastic scattering measurements.\\

{\bf Acknowledgment} We thank L. Daemon for help in acquiring 
X-ray diffraction data for our samples.   This work was performed under the auspices of the U.S. 
Department of Energy under Contract DE-AC52-06NA25396. Author DJS is
supported by the DOE Office of Science Graduate Fellowship Program, 
made possible in part by the American Recovery and Reinvestment Act of
2009, administered by ORISE-ORAU under Contract DE-AC05-06OR23100.

\begin{figure}[h] 
\centering
\includegraphics[width=5.2in,angle=0]{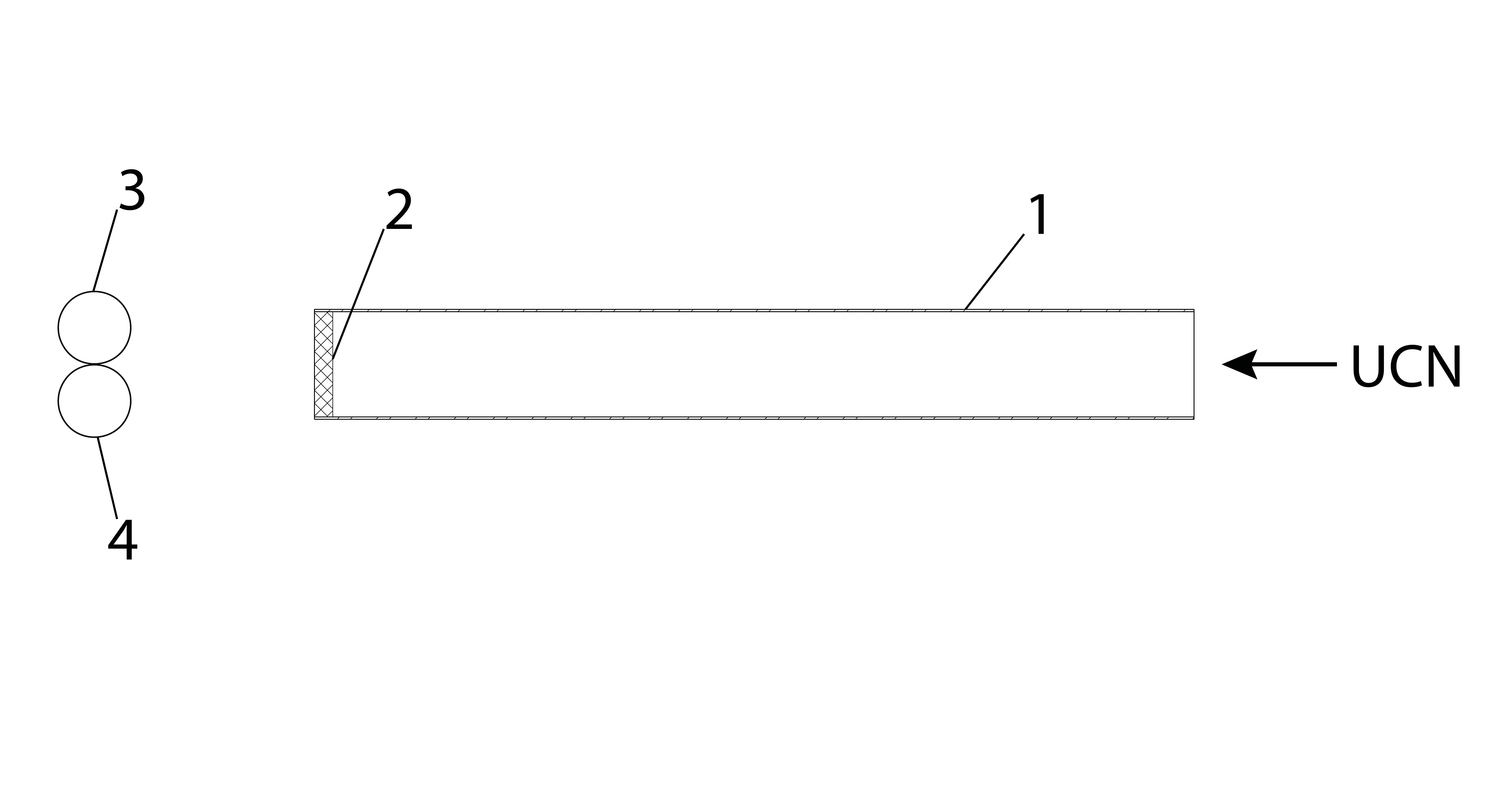}
\caption{Scheme of a section of the neutron guide and the drift-tube detectors placement 
for measuring upscattering of ultracold neutrons: 1) is the neuron guide, 2) is the sample, and 3) and 4) are the neutron detectors.}
\label{Fig1}
\end{figure}

\begin{figure}[h] 
\centering
\includegraphics[width=5.2in,angle=0]{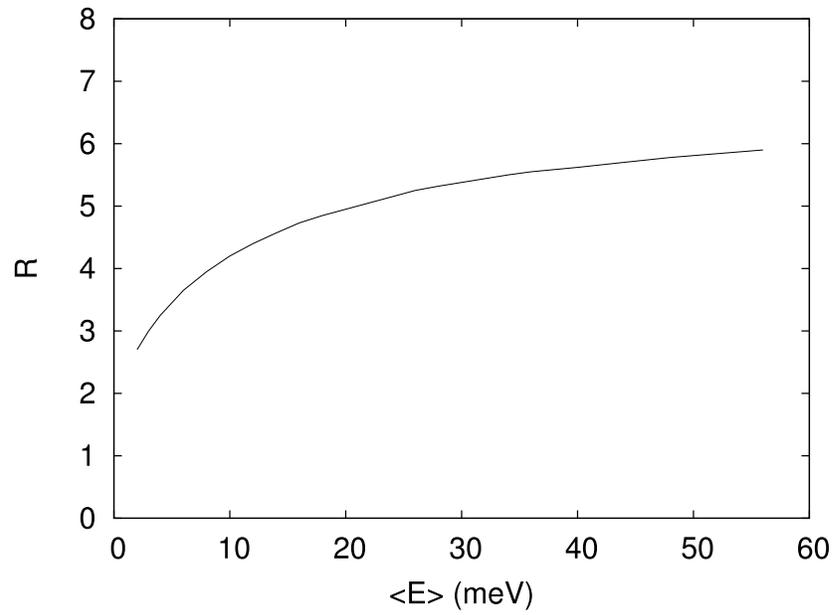}
\caption{ The calculated ratio R of the LANL drift-tubes (with the 1.8 and 0.2 bar of He-3
partial pressure) 
 efficiencies versus the average energy $\langle$E$\rangle$ of the Maxwellian scattered flux.  It would give the 
result for $\langle$E$\rangle$ from measured R if the assumption of
the Maxwellian shape for the neutron flux is valid. As explained in text we do not use this assumtion in analyzing our data. }   
\label{Fig2}
\end{figure}

\begin{figure}[b] 
\centering
\includegraphics[width=5.2in,angle=0]{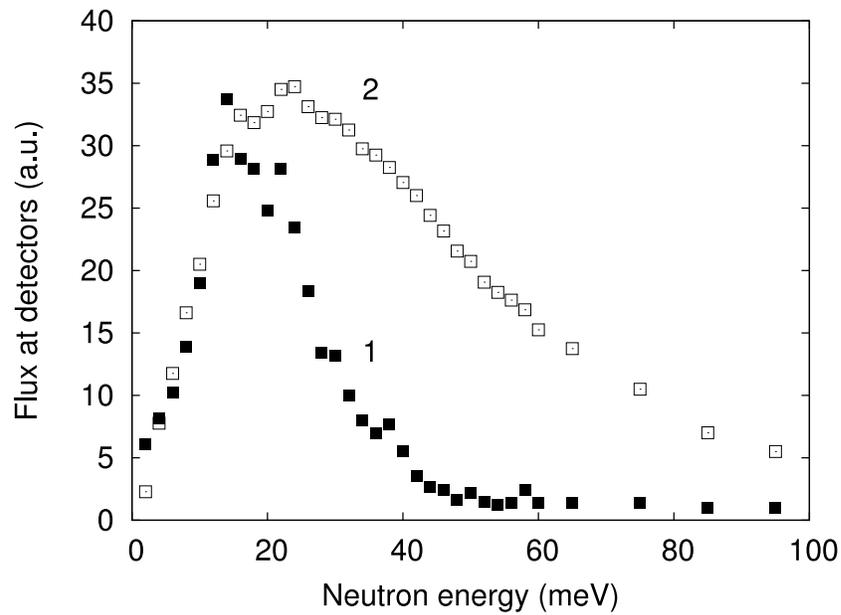}
\caption{ The calculated energy spectra of the neutron flux scattered from  
the  PE sample, 500 $\mu$m thick, E$_{in}$=100 neV (curve 1), and of the fully moderated flux (curve 2) after 
"filtering" this flux through the 6-mm polyethylene slab.  
The spectra are obtained by MCNP modeling with the 293K S($\alpha,\beta$)
scattering law. 
The shape of spectrum `1' is not Maxwellian, it has  
the average energy value of 26$\pm 3$ meV. Spectrum 2 is a  Maxwellian
with the average energy of 53$\pm 4$ meV.}
\label{Fig3}
\end{figure}

\begin{figure}[b] 
\centering
\includegraphics[width=5.2in,angle=0]{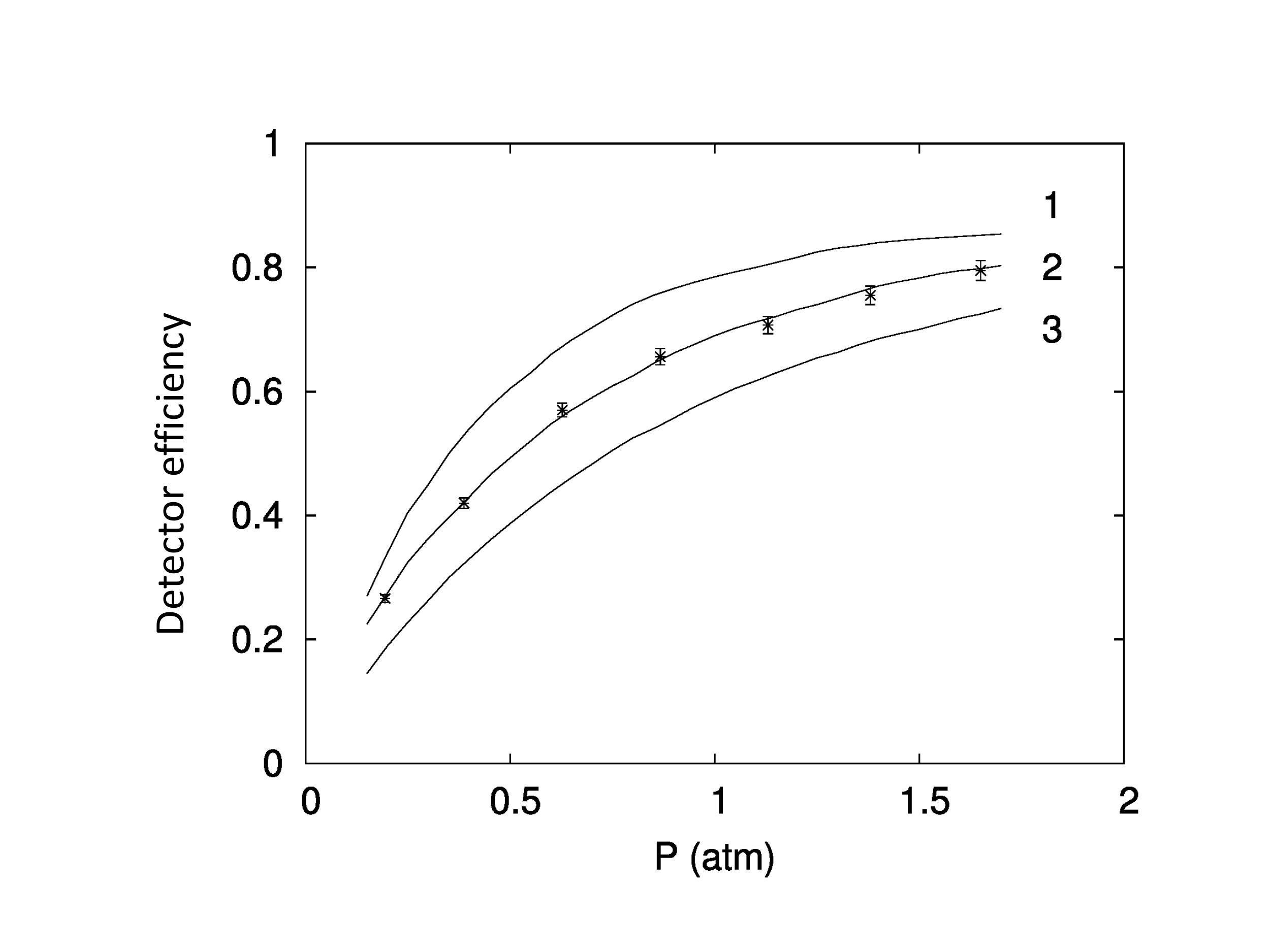}
\caption{Reanalysis of the Stoika and Strelkov \cite{Str78} PE experiment. 
Experimental data \cite{Str78} for  their 
neutron  detector efficiencies as a fuction of the He-3 pressure  
are shown as crosses. Curves 1 and 2 $-$  show our MCNP
modeling of 
their detector efficiencies, which we performed using
the PE scattering law S($\alpha,\beta$) kernels at 77K and 293K. 
Curve 1 corresponds to non-Maxwellian upscattered 
spectrum with $\langle$E$\rangle$=11 meV, for 
getting which  we resorted to the 77K kernel. 
Curve 2 corresponds to the 293K spectrum with $\langle$E$\rangle$=26 meV.
 Curve 3 for the 293K Maxwellian 
spectrum is shown, following \cite{Str78}, for comparison. 
From agreement of data \cite{Str78} 
with the curve 2 we 
deduce  the average energy of 26 meV for their PE experiment, 
which is the same as  
the result of our PMP experiment.}
\label{Fig4}
\end{figure}

\end{document}